\documentclass[aps,prl,floats,twocolumn,superscriptaddress,amssymb,amsmath,showpacs]{revtex4}

\usepackage{graphicx}
\usepackage{epstopdf}
\begin{document}
\title{Casimir forces, surface fluctuations, and thinning of 
superfluid films}

\author{Roya Zandi}
\affiliation{Department of Chemistry and Biochemistry, UCLA, Box
951569, Los Angeles, CA 90095-1569}
\affiliation{Department of Physics, Massachusetts Institute of
Technology, Cambridge, MA 02139, USA}

\author{Joseph Rudnick}
\affiliation{Department of Physics, UCLA, Box 951547, Los Angeles,
CA 90095-1547}

\author{Mehran Kardar}
\affiliation{Department of Physics, Massachusetts Institute of
Technology, Cambridge, MA 02139, USA}

\date{\today}

\begin{abstract}

Recent experiments on the wetting of $^{4}$He have shown that the film
becomes thinner at the $\lambda$ transition, and in the superfluid
phase.  The difference in thickness above and below the transition has
been attributed to a Casimir interaction which is a consequence of a
broken continuous symmetry in the bulk superfluid.  However, the
observed thinning of the film is larger than can be accounted by this
Casimir force.  We show that surface fluctuations give rise to an
additional force, similar in form but larger in magnitude, which may
explain the observations.

\end{abstract}

\pacs{46.32.+x, 82.35.Rs, 87.15.La, 85.35.Kt, 07.10.Cm} \maketitle

%

Quantum fluctuations of the electromagnetic field between two
conducting plates in vacuum result in long-ranged attractive
interactions.  This effect which has only recently experimentally verified by a series of
high precision measurements\cite{Casimir},
was first predicted by Casimir in 1948\cite{CasimirX}.
Thirty years later, Fisher and de Gennes noted that the confinement of 
thermal fluctuations in fluids 
leads to similar long-ranged forces\cite{fisher}.  
Quite generally, geometric restrictions on a fluctuating field constrain 
the normal modes of fluctuations and result in fluctuation-induced or
Casimir forces.  These forces are controlled by correlations in the fluid; when the correlations are long-ranged, corresponding to massless fields, Casimir forces decay with distance as a simple power law\cite{lifshitz,indekeu,ajdari,li,krech,krech2,golestan,most}.  The overall strength of a Casimir interaction is typically universal.  That is, it depends on symmetries of the fluctuating field, and on boundary conditions, but not on microscopic details.

\begin{figure}
\includegraphics[width=0.8\columnwidth]{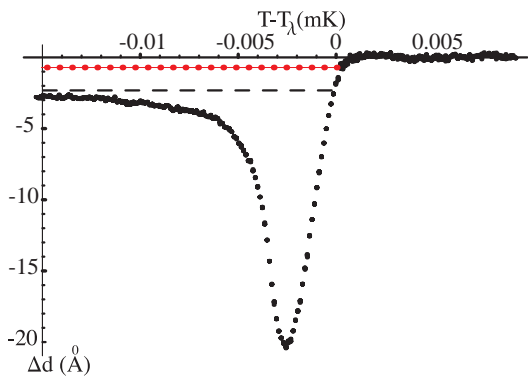}
\caption{ The thinning of the Helium film $\Delta d$ versus temperature,
based on the data reported in Refs.~\cite{garcia,note}. 
The dotted line illustrates the expected thinning of the film due to a
Casimir force resulting from bulk Goldstone modes.  It also coincides with
the magnitude of the thinning of the film at the transition temperature as
determined in Ref.~\cite{krech2}.  The dashed line shows the expected
change in thickness due to {\em both} bulk Goldstone modes and surface fluctuations.}
\label{fig:garcia}
\end{figure}

An important example of a Casimir force associated with thermal
fluctuations in a condensed matter system is found in $^{4}$He
films at and near the superfluid phase transition\cite{garcia}.  The
finite thickness, $d$, of the film constrains the fluctuations
of the superfluid order parameter, which then mediate a Casimir force.
Experimental demonstration of this force was reported recently by 
Garcia and Chan\cite{garcia} (GC). To produce films of various thicknesses, a stack
of copper electrodes were suspended at different heights above bulk
liquid helium.  The thickness of the wetting layer on each electrode
as a function of temperature was monitored to gauge the strength of
interactions with the substrate.  Figure~\ref{fig:garcia} shows the change in the
film thickness $\Delta d =d-d_{0}$, as a function of reduced
temperature $t=T-T_{\lambda}$, near the superfluid transition point for
the capacitor labeled ``Cap 1'' in Ref.~\cite{garcia}.  The quantity
$d_{0}$ is the thickness of the film well above the
$\lambda$-point.  As shown in the figure, there is a perceptible
decrease in the thickness of the film at the transition point $t=0$,
followed by a substantial drop below the transition.  The thinning of
the film for $t\geq 0$ quantitatively confirms the theoretical predictions of
attractive Casimir interactions between parallel surfaces in the
presence of Dirichelet boundary conditions\cite{krech2}.  At this
point, no theoretical explanation has been put forth for the
relatively substantial dip in the film thickness at temperatures close
to and below the bulk transition temperature.  In fact, the
theoretical magnitude of the Casimir force at $t=0$ is 50 times less
than required to give rise to the observed maximum thinning
\cite{balibar}.

Further below the transition point, the superfluid film partially recovers its
thickness, but still remains noticeably thinner than in the normal fluid.  
The superfluid state, in which a continuous symmetry is broken, 
supports Goldstone modes that are not present in the normal phase.
It is thus reasonable to expect that these long wavelength modes are
responsible for the differences in thickness on the two sides of the 
transition point.  However, the magnitude of the fluctuation-induced force
associated with the Goldstone modes in the bulk of the film is too small to
account for the observed reduction in thickness in the superfluid 
region~\cite{balibar,garciathesis} (away from criticality).  The dotted line in 
Fig.~\ref{fig:garcia} indicates the expected thinning of the film due to the Casimir
force resulting from the bulk Goldstone modes.

Here, we explore the role of surface fluctuations on the
thinning of a superfluid film.  In particular, we investigate the
fluctuation-induced force generated by the flow field in a superfluid film
that accompanies undulations in the fluid-vapor interface. 
While it may initially appear that surface effects should be subdominant
to those from the bulk, the flow fields (and hence kinetic energy) associated
with surface deformations are actually quite constrained by
the film thickness, $d$. The dependence of undulation energies on thickness leads to a
contribution to the free energy that is proportional to $1/d^2$.
This leads to a force favoring thinner films, with the same form as that arising from
the bulk Goldstone modes, and an amplitude that is almost twice as large.
The dashed line in Fig.~\ref{fig:garcia} shows the expected change in the 
thickness of the film due to the combined  influences of surface and bulk
fluctuation-induced forces for a film of thickness 423~\AA.
As is clear from the figure, the net effect  of  ``bulk'' Goldstone modes
and surface capillary modes is sufficient
to explain the diminished breadth of films in the superfluid phase.

We would like to emphasize that a theoretical calculation of forces
induced by surface fluctuations in liquid films is not without 
precedent~\cite{lifshitz,cole}.
However, we are unaware of any instance in which such interactions have 
been detected prior to the experiments of Garcia and Chan. 
The key point is that in most cases, fluctuation-induced interactions
(due to bulk phonons or surface modes) appear only as a small correction
to the larger van der Waals forces.
Traversing the superfluid transition, however, provides an instance in
which the dominant atomic forces are unchanged, while the long wave-length
modes due to continuous symmetry breaking can be switched on or off.
Thus the {\em change in thickness} of the wetting film provides an ideal 
probe of the interactions generated by the Goldstone modes.

Deep in the superfluid state, the magnitude of the complex order parameter 
is fixed, but its phase $\phi$ can vary with relatively small energy cost.
Spatial variations of $\phi$ are accompanied by the flow of the superfluid
component. This connection underlies the two-fluid model of
superfluidity\cite{landau}, which leads to the m\'{e}lange of
acoustic and wavelike excitations supported by this system,
including first, second, third and fourth sound~\cite{putterman,tilley}.
For our purposes, we decompose the variations of $\phi$ into components
arising from bulk and surface modes.
In terms of the wave-vector ${\bf k}\equiv(k_x,k_y,k_z=n \pi/d)$, the bulk modes have the form
\begin{equation}
\phi_b({\bf k}) = A_{\bf k}~e^{i k_x x +i k_y y} \cos \left(n\pi z\over d\right). \label{eq:phibulk}
\end{equation} 
The cosine in Eq.~(\ref{eq:phibulk}) guarantees that $\partial
\phi/\partial z$ is equal to zero at $z=0$ and $z=d$, so that the flow
field associated with phase fluctuations does not cause a displacement
of either of the two interfaces.  
The above mode is accompanied by a 
superfluid velocity ${\bf v}_s=\hbar \nabla\phi_b/m$, and a corresponding kinetic 
energy $(\rho_s/2)\int d^3x v_s^2=(\hbar^2 \rho_s V/4m^2)k^2 |A_{\bf k}|^2$, 
where $\rho_s$ is the superfluid density, and $m$ is the mass of a helium atom.

Integrating over all decompositions of the phase $\phi$ according to Eq.~(\ref{eq:phibulk}),
we obtain the free energy associated with this set of bulk Goldstone modes as
\begin{equation}
{\cal F}_b = \frac{k_{B}T}{2}A \int \frac{d^{2}q }{( 2 \pi)^{2}}\ 
\sum_{n} \ln \left[\left( q^{2}+ \left(n \pi\over d\right)^{2}\right)\right] + \mathcal{F}^{\prime}.
\label{free2}
\end{equation}
Here, $A$ is the total area of the film, and $\mathcal{F}^{\prime}$ corresponds to contributions 
that do not generate a non-trivial dependence on $d$.
Indeed, the sum over $n$ can be performed by standard contour integration techniques\cite{arfken};
while the dominant term is an extensive contribution to the free energy,
there is an important correction that scales as $1/d^2$.  
The  latter is the fluctuation-induced interaction, which leads
to a Casimir force per unit area, of~\cite{li}
 \begin{equation}
F_{\rm bulk}=-\frac{k_{B}T}{8 \pi} \frac{\zeta(3)}{d^{3}} .
\label{casforce}
\end{equation}
 
The thickness of the adsorbed helium film is determined by the competition
of several factors, notably the loss of gravitational potential energy, and attractions to the substrate.
The former can be calculated simply from the height difference $h$ between 
the adsorbing plate and the bulk liquid, while the latter is due to the van der Waals
interactions with the substrate\cite{garcia,chen}.
Fluctuation induced forces, as in Eq.~(\ref{casforce}), provide an additional component.
The film thickness $d$ is thus determined by the force balance equation
\begin{equation}
\label{chan}
 \frac{\gamma}{d^{3}}  \left(1+\frac{d}{d_{1/2}}\right)^{-1} + 
\frac{k_{B}T v}{d^{3}} \vartheta=mgh .
\end{equation}
The first term on the left hand side is the van der Waals interaction,
with a leading behavior of $\gamma/d^3$ with 
$\gamma =2600^\circ$K~\AA$^{3}$ for a film of $^{4}$He on Cu.  
Retardation effects due to the finite speed of light are significant for $d$
of the order of  $d_{1/2}=193$~\AA, necessitating the inclusion of the correction term.
The second term on the left hand side is the Casimir force, which has the
same leading behavior as the van der Waals term, with a magnitude set by
$v=45.81$~\AA$^{3}$/atom and the dimensionless amplitude $\vartheta$\cite{Qcorrections}.

Unlike the van der Waals interaction, the parameter $\vartheta$ is expected to
change rapidly at the superfluid transition.
It is zero in the normal phase, while fluctuations of the superfluid order parameter
lead to an interaction that is  `attractive' in the sense of favoring a thinner film.
On approaching the transition from the normal liquid, the amplitude $\vartheta$
is a scaling function of  $d~(T-T_\lambda)^\nu$ ($\nu$ is the exponent for the
divergence of the correlation length) which was obtained in
a two-loop Renormalization Group calculation by Krech\cite{krech}. 
The predicted thinning of the nearly superfluid film closely tracks the observations,
e.g. as in the portion of Fig.~\ref{fig:garcia} for $T\geq T_\lambda$.
Currently, there are no calculations that reproduce the large dip in thickness
for $T\leq T_\lambda$\cite{GWilliams}.
Well below the transition, in the superfluid phase, $\vartheta$ is expected to be
a universal constant, such as in Eq.~(\ref{casforce}).
The maximum amplitude calculated by Krech~\cite{krech2} is coincidentally rather close to
the value given in Eq.~(\ref{casforce}) coming from the Goldstone modes in the bulk of the film.
(Monte Carlo simulations with periodic boundary conditions in three dimensions 
yield a value of the critical amplitude which is roughly twice larger\cite{KrechMC}.)
Thus, if these modes were the  only cause for the thinning of the film in the
superfluid phase,  the height of the film would be roughly the
same at the critical point and well below the $\lambda$-point.  
As  indicated in Fig.~\ref{fig:garcia}, this is not the case.
More precisely, if $h= 0.228$~cm ($mgh = 10.76~\mu^\circ$K) the equilibrium film thickness 
can be deduced from Eq.~(\ref{chan}) to be $d=423$~\AA\
above the transition temperature.  This is the thickness of
the film for ``Cap.~1'' in the experiment of GC.
In the case of the bulk Goldstone modes, $\vartheta = -\zeta(3) / 8 \pi$, which is
too small to explain the reduction in thickness below the transition.  
A straightforward calculation indicates that the thinning
of the film is consistent with a force which is at least three times
as large as the one at the bulk critical point ($T=T_{\lambda}$).

To resolve the discrepancy with the experimental observation, we
now resort to the effect of surface fluctuations in the superfluid\cite{Qcorrections}.
According to Dzyaloshinkskii, Lifshitz, and Pitaevskii (DLP), surface
fluctuations in fluids also act as a source for Casimir forces\cite{lifshitz}.  
However, in the case of thin liquid films, viscous damping effectively clamps the fluid, 
and there is no indication that such forces play any role in the thinning of the 
helium films.

But when the film is in the superfluid state, there is no viscosity opposing the flow
fields that accompany surface deformations.
To quantify the effect of surface fluctuations, we consider a film of thickness
\begin{equation}
d(\vec{R},t)=d+ h_{\vec{q}}(t)~e^{i\vec{q} \cdot \vec{R}} ,
\label{d(t)}
\end{equation} 
where $\vec{q} = \hat{x} q_x + \hat{y} q_y$ is the two-dimensional wave
vector of the surface distortion, and $\vec{R}$ is a
two-dimensional position vector.  
This deformation is accompanied by a distortion in the phase of the superfluid
order parameter of 
\begin{equation} 
\phi_s(\vec{q}\,) =\frac{m}{\hbar}~\dot{h}_{\vec{q}}~ 
\frac{\cosh(qz)}{q\sinh(qd)}~e^{i\vec{q}~ \cdot \vec{R}} ~~.
\label{eq:phis} \end{equation} 
The form of $\phi_s$ is chosen such that the vertical velocity $v_z=(\hbar/m)\partial_z\phi_s$,
is zero at the substrate ($z=0$), and coincides with the motion of the liquid-vapor interface
at $z=d$.
Variations of $\phi_s$ along the $z$ direction are exponential in $qz$, with
$q=\pm\sqrt{q_x^2+q_y^2}$. This choice ensures that $\nabla^2 \phi_s =0$,
such that the kinetic energy is minimal, and that
there are no couplings to the bulk modes considered earlier.
(We note that the velocity potential in Eq.~\ref{eq:phis} reproduces the flow 
field associated with the third sound mode~\cite{atkins}.)

The kinetic energy in the flow set up by $\phi_s$ is
\begin{equation} 
\frac{\rho_s}{2}\int d^3x \left(\frac{\hbar}{m}\nabla\phi_s\right)^2=
A~\frac{\rho_s}{2}~|\dot{h}_{\vec{q}}|^{2}~\frac{\coth(qd)}{q}.
\label{eq:newkin1}
\end{equation} 
Surface deformations are also accompanied by a potential energy, and surface
tension cost. Including these contributions, and summing over all surface modes,
leads to a Hamiltonian
\begin{equation}
{\cal H}_s=\sum_{\vec{q}} \left[ |\pi_{\vec q}|^{2} \frac{q \tanh( qd)}{ 2 \rho_{s}} +
|h_{\vec q}|^{2} \left(\frac{\rho f}{2}+\frac{\sigma}{2} q^{2}\right) \right].
\label{eq:toten1}
\end{equation}
In the above equation, we have introduced the conjugate momentum
\begin{equation} 
\pi_{\vec q} =  \frac{\rho_s}{q \tanh qd}~\dot{h}_{\vec{q}}~;
\label{eq:pi1} 
\end{equation} 
in the potential energy part $\rho$ is the mass density,  $\sigma$ is
the surface tension, and $f$ is the net force on the surface particles.

The classical partition function\cite{Qcorrections} associated with the surface modes is easily
obtained by integrating  over $\pi_{\vec{q}}$ and $h_{\vec{q}}$, leading to 
a contribution to the free energy of the film of
\begin{eqnarray}
{\cal F}_s=\frac{1}{2}k_{B}T \sum_{\vec{q}} \left[ \ln \left(\frac{q \tanh (qd)}{\rho_{s}}\right)
+ \ln ( \rho f +\sigma q^{2}) \right] .
\label{eq:newfree1}
\end{eqnarray}
Note that the contribution from the potential energy (the integral over $h_{\vec{q}}$)
is present both above and below the transition, while the kinetic energy term
(from integration over $\pi_{\vec{q}}$) exists only in the superfluid phase.
In fact, it is only the latter that explicitly depends on the thickness of the film
through the factor of $\tanh(qd)$:
the $d$ dependence of the free energy comes entirely from 
$ (k_BT /2)\sum_{\vec{q}} \ln\left[ \tanh(qd)\right] $,
and is independent of various material dependent parameters appearing 
in Eq.~(\ref{eq:newfree1})\cite{third-sound}.
Taking a derivative with respect to $d$, we obtain the
Casimir force resulting from the superfluid flow field induced by
surface fluctuations, as
\begin{equation} 
F_{\rm surface} = -
\frac{7}{4}\frac{k_BT }{8 \pi }\frac{\zeta(3)}{d^3}~,
\label{eq:Casforces} 
\end{equation} 
which is nearly twice as large as the ``bulk'' Goldstone mode force in Eq.~(\ref{casforce}).

The total Casimir force per unit area due to the surface fluctuations
and Goldstone modes is
\begin{equation}
\label{energy3}
F_{\rm casimir} = -\frac{11}{4}\frac{k_BT }{8 \pi }\frac{\zeta(3)}{d^3}
\approx -0.15\frac {k_BT }{d^{3}}.
\end{equation}
Using Eqs.~(\ref{chan}) and (\ref{energy3}), we find 
$\Delta d \approx 2.2$\AA \ at temperatures slightly below
$T_{\lambda}$,
for a film whose thickness above the transition is $d=423$\AA. 
The dashed line in Fig.~\ref{fig:garcia} illustrates this expected change in
thickness of the film, which is very close to that observed in the experiments of GC.
It is important to keep in mind that the interpretation of wetting experiments on
$^{4}$He is complicated by the presence of microscopic scratches, dust
particles and surface roughness.  All these have a noticeable effect
on the estimation of the thickness of the film~\cite{garcia,krech2}.

In conclusion, experiments on helium films adsorbed on copper surfaces
provide a powerful tool for probing
fluctuation-induced forces, as the effects due to the
superfluid order parameter are turned on at the transition temperature,
while the atomic forces are essentially unchanged.
Fluctuations in the phase of the order parameter are an evident candidate,
but can only partially account for the observed thinning of the superfluid film.  
We suggest that surface fluctuations provide additional  interactions.
In particular, in the superfluid phase the motion of the liquid-vapor interface 
sets up a superfluid velocity field that extends throughout the film.
The corresponding fluctuation-induced force has exactly the same (material independent)
form as that from the bulk phase fluctuations, but with an amplitude that is
nearly twice as large.
Considering the statistical and systematic errors in the experiments, 
it is reasonable to conclude that the combination 
of the bulk and surface fluctuation-induced forces is sufficient to explain
the excess thinning of the film in the superfluid region.  
The relatively large error bars on the helium wetting experiments
\cite{garcia,balibar,ueno}, do not mitigate our results, which relate
to the difference in film thickness above and below the $\lambda$
transition.  One important extension of this work will be to find the
effect of surface fluctuations on the Casimir force at---and
especially immediately below---the superfluid transition.

The authors would like to acknowledge helpful discussions with A.
Gopinathan and G. A. Williams, and thank M.~H.~W. Chan and R. Garcia for the
data displayed in Fig.~\ref{fig:garcia}.  RZ acknowledges support
from the UC President's Postdoctrol Fellowship program.
This work was  supported by NSF grant DMR-01-18213 (MK).


\begin{thebibliography}{99}
\bibitem{Qcorrections}
Since we emphasize the importance of retardation corrections for the van der Waals term,
the reader may wonder about related quantum corrections to the Casimir term.
The relevance of quantum corrections is determined by $\hbar v / (k_{B}T d)$, 
where $v$ is the speed of sound in the helium film~\cite{schwinger}.  
Given that $v$ is of order of 100 meters per second, and $d$ is about 400~\AA, 
we find $\hbar v / (k_{B}T d) \sim 10^{-3}$, which is insignificant.
Quantum corrections become importance for films of thickness  $d\sim400$~\AA,
only at temperatures of around 1~m$^\circ$K.
\bibitem{GWilliams}
A calculation by G. Williams\cite{gw} based on the action of vortex loops,
but neglecting the Goldstone modes, in the superfluid phase, produces a fluctuation
induced force with an amplitude $\vartheta$ that closely mirrors the results of Krech\cite{krech}
in the normal phase.
\bibitem{third-sound}
From the Hamiltonian in Eq.~(\ref{eq:toten1}) we can extract the velocity of the surface modes.
The final result is very close to the form given in Ref.~\cite{atkins} for that of third sound, 
except in the placements of $\rho_s$ and $\rho$. The distinction disappears deep in the superfluid
(the regime of interest to us) where $\rho_s\approx\rho$. In any case, these parameters
do not appear in the final result of Eq.~(\ref{eq:Casforces}) which is robust.
\end{thebibliography}
\end{document}